\documentclass[12pt]{article}
\usepackage{epsf}
\def\beq{\begin{equation}}
\def\eeq{\end{equation}}
\def\bea{\begin{eqnarray}}
\def\eea{\end{eqnarray}}
\def\ve{\vert}
\def\vel{\left|}
\def\ver{\right|}
\def\nnb{\nonumber}
\def\ga{\left(}
\def\dr{\right)}

\def\rar{\rightarrow}
\def\nnb{\nonumber}
\def\la{\langle}
\def\ra{\rangle}
\def\ba{\begin{array}}
\def\ea{\end{array}}

\title{ {\small {\bf RARE RADIATIVE 
$B \rar \tau^+ \tau^- \gamma$ DECAY IN THE TWO HIGGS DOUBLET MODEL} } }
\author{ {\small E. O. \.{I}LTAN and  G. TURAN}\\
{\small  Physics Department, Middle East Technical University} \\
{\small 06531 Ankara, Turkey }}
\date{}
\begin{document}
\setlength{\baselineskip}{24pt}
\maketitle
\setlength{\baselineskip}{7mm}
\begin{abstract}
The radiative $B \rar \tau^+ \tau^- \gamma$ decay is investigated in the 
framework of the two Higgs doublet model . The dependence of the differential 
branching ratio on the photon energy and the branching ratio on the two Higgs 
doublet model parameters, $m_{H^\pm}$ and $\tan \beta $, are studied. It is 
shown that there is an enhancement in the predictions of  the  two Higgs 
doublet model compared to the Standard model  case. We also observe that
contributions of neutral Higgs bosons to the decay are sizable when 
$\tan\beta $ is large.
\end{abstract}
\thispagestyle{empty}
\newpage
\setcounter{page}{1}
\section{Introduction}
Rare B-meson decays are one of the important research areas to test the 
theoretical models  and make estimations about their free parameters. 
In the Standard model (SM) they are induced by flavour changing neutral 
currents (FCNC) at the loop level. This ensures a precise determination 
of the fundamental parameters of the SM, such as Cabbibo-Kabayashi-Maskawa 
(CKM) matrix elements, leptonic decay constants, etc. In addition, the 
studies on rare B-meson decays give powerfull clues about the existence of 
model beyond the SM, such as two Higgs doublet model 
(2HDM), minimal supersymmetric extension of the SM (MSSM) \cite{Hewet}, etc. 
Among rare B-decays, $B \rar \ell^+ \ell^- \gamma $ decays are of special 
interest due to their cleanliness and  sensitivity to the new physics. They 
have been investigated in the framework of  the SM in \cite{Eilam1,Aliev1} for
$\ell =e,\mu $ and in \cite{Aliev2} for $\ell =\tau$. The theoretical results 
given in \cite{Aliev1} and
\cite{Aliev2} are BR($B_{s}\rightarrow e^+ e^- \gamma )=2.35 \times 10^{-9}$,  
BR($B_{s}\rightarrow \mu^+ \mu^- \gamma )=1.9 \times 10^{-9}$ and 
BR($B_{s}\rightarrow \tau^+ \tau^- \gamma )=9.54 \times 10^{-9}$, respectively.
These decays get negligible contributions from the diagrams, where photon is 
radiated from any charged  internal line due to the fact that they will
have a factor $m_{b}^{2}/M_{W}^{2}$ in the Wilson coefficients. When photon is 
radiated from the final charged leptons, the contribution is proportional to 
the lepton mass $m_{\ell }$. Therefore,  for $\ell =e, \mu $ case, it is 
negligible; however for $\ell =\tau$ it gives a considerable contribution to 
the amplitude. In the 2HDM, there is a part coming  from exchanging  neutral 
Higgs bosons and in contrast to  $B \rar \ell^+ \ell^- \gamma $ ($\ell =e,\mu $) 
decays , we could expect that they significantly contribute for 
$ B_{s}\rightarrow \tau^{+} \tau^{-} \gamma $ decays. Therefore,  in this 
work we study the   $ B_{s}\rightarrow \tau^{+} \tau^{-} \gamma $ process in 
the framework of the 2HDM (Model I and II).

2HDM is one of the simplest extensions of the SM, obtained by the addition of 
a second Higgs doublet. In this model,  there are one physical charged Higgs 
scalar, 
two neutral Higgs scalars and one neutral Higgs pseudoscalar. The Yukawa 
lagrangian causes that the model possesses  tree-level FC couplings 
of the neutral Higgs particles. To avoid such terms, it is proposed an ad hoc 
discrete symmetry \cite{Glashow} on the 2HDM potential and the Yukawa 
interaction. As a result, it appears two different choices for how to couple 
the quarks to the two Higgs doublets: In the first choice (Model I), the quarks 
do not couple to the first Higgs doublet, but couple to the second one.  In 
the second choice, (Model II), the first Higgs doublet couples only to 
down -type quarks and the second one  to only up-type quarks.

The paper is organized as follows: In sec.2, we present the theoretical 
framework for the $ B_{s}\rightarrow \tau^{+} \tau^{-} \gamma $ decay 
and describe some details of its decay rate calculation . We give  a 
numerical analysis and discussion of our results in sec.3. Appendices 
contain a list of  the operators and the Wilson coefficients, as well 
as some  relevant formula about the long distance contributions.

\section{  $ B_{s}\rightarrow \tau^{+} \tau^{-} \gamma $ decay in the 
framework of the 2HDM }
The exclusive decay $ B_{s}\rightarrow \tau^{+} \tau^{-} \gamma $ can 
be obtained from the inclusive one  $ b\rightarrow s \tau^{+} \tau^{-} \gamma $. 
In order to calculate the relevant physical quantities for  the decay  
$ b\rightarrow s \tau^{+} \tau^{-} \gamma $, we start with the QCD corrected 
amplitude for the process $ b\rightarrow s \tau^{+} \tau^{-}  $. At this 
stage, the effective Hamiltonian is obtained by matching the full theory 
with the effective low energy one at the high scale $\mu$. The Wilson 
coefficients are evaluated from $\mu$ down to the lower scale 
$\mu \sim O(m_{b})$ using the renormalization group equation (RGE). The 
effective Hamiltonian in the 2HDM for the process 
$b\rightarrow s \tau^+\tau^-$ is \cite{Dai}
\beq
{\cal H}=\frac{-4\, G_F}{ \sqrt{2}} V_{tb} V_{ts}^*
\Bigg{\{}\sum_{ i =1}^{10} C_{i}(\mu ) O_{i}(\mu)+
\sum_{ i =1}^{10} C_{Q_{i}}(\mu )Q_{i}(\mu) \Bigg{\}}
\label{H1}
\eeq
In this equation $O_{i}$ are current-current $(i=1,2)$, penguin $(i=1,..,6)$, 
magnetic penguin $(i=7,8)$ and semileptonic $(i=9,10)$ operators . The 
additional operators $Q_{i}, (i=1,..,10)$ are due to the neutral Higgs boson 
exchange diagrams, which give considerable contributions in the case that the 
lepton pair is $\tau^{+}\tau^{-}$ \cite{Dai}. 
$C_{i}(\mu )$ and $C_{Q_{i}}(\mu )$ are Wilson coefficients renormalized at 
the scale $\mu$. All these operators and the Wilson coefficients, together 
with their initial values calculated at $\mu=m_W$ in the SM and also the 
additional coefficients coming from the new Higgs scalars are presented in 
Appendices A and B. The QCD corrected amplitude for the inclusive 
$b\rightarrow s \tau^{+}\tau^{-}$ decay in the 2HDM (Model I or II) is
\bea
{\cal M} & = & \frac{\alpha G_F}{ \sqrt{2}\, \pi} V_{tb} V_{ts}^*
\Bigg{\{} C_9^{eff} (\bar s \gamma_\mu P_L b) \, \bar \tau \gamma_\mu \tau +
C_{10} ( \bar s \gamma_\mu P_L b) \, \bar \tau \gamma_\mu \gamma_5 \tau    
\nnb \\
& -& 2 C_7 \frac{m_b}{p^2} (\bar s i \sigma_{\mu \nu} p_\nu P_R b) \bar \tau 
\gamma_\mu \tau 
+ C_{Q_{1}}(\bar s \gamma_\mu P_R b) \bar \tau  \tau +C_{Q_{2}}
(\bar s \gamma_\mu P_R b) \bar \tau \gamma_5 \tau \Bigg{\}}~.
\eea
where $P_{L,R}=(1\mp \gamma_5)/2$ , $p$ is the momentum transfer and 
$V_{ij}$'s are the corresponding elements of the CKM matrix.  

In order to obtain the matrix element for 
$b \rightarrow s \tau^{+}\tau^{-} \gamma $ decay, a photon line should be 
attached to any charged internal or external line. The contributions coming 
from the attachement of  photon to any internal line are suppressed and we 
neglect them in the following analysis. We now start with the case in which  
a photon is attached  to the initial quark lines. The corresponding matrix 
element for the $B_s \rar \tau^+ \tau^- \gamma$ decay is 
\bea
{\cal M}_1 &=& \la \gamma \ve {\cal M}\ve B \ra =                         
\frac{\alpha G_F}{2 \sqrt{2} \, \pi} V_{tb} V_{ts}^* \nnb   \Bigg{\{} C_9^{eff}  
\bar \tau \gamma_\mu \tau \la \gamma \ve \bar s \gamma_\mu (1-\gamma_5) b \ve  
B \ra  + C_{10} \, \bar \tau \gamma_\mu \gamma_5 \tau \la \gamma \ve \bar s 
\gamma_\mu(1-\gamma_5) b \ve B\ra \nnb \\                  
&-&  2 C_7 \frac{m_b}{p^2} \la \gamma \ve \bar s i \sigma_{\mu \nu} p_{\nu} (1+ 
\gamma_5) b \ve B \ra \bar \tau \gamma_\mu \tau  
+ C_{Q_{1}} \bar \tau  \tau \la \gamma \ve \bar s (1+\gamma_5) b \ve B\ra + 
C_{Q_{2}} \bar \tau  \gamma_5\tau \la \gamma \ve \bar s (1+\gamma_5) 
b \ve B\ra  \Bigg{\}} \nnb \\ & &  \label{M1eq1}                         
\eea 
 
These matrix elements can be written in terms of the two independent, gauge
invariant, parity conserving and parity violating form factors 
\cite{Aliev1,Eilam2}:
\bea
\la \gamma \ve \bar s \gamma_\mu ( 1\mp \gamma_5) b \ve B \ra &=&
\frac{e}{m_B^2} \Bigg{\{}                                                      
\epsilon_{\mu \alpha \beta \sigma} \epsilon_\alpha^* p_\beta q_\sigma   
\, g(p^2)  \pm ~i \left[ \epsilon_\mu^* (p q ) - (\epsilon^* p) q_\mu\right] 
\, f(p^2) \Bigg{\}}~, \label{ff1} 
\eea
and
\bea                                           
\la \gamma \ve \bar s i \sigma_{\mu \nu} p_\nu (1\mp \gamma_5) b \ve
B \ra &=& \frac{e}{m_B^2} \Bigg{\{} \epsilon_{\mu \alpha \beta \sigma}
\epsilon_\alpha^*
p_\beta q_\sigma \, g_1(p^2)              
\mp ~ i \left[ \epsilon_\mu^* (p q) - (\epsilon^* p ) q_\mu \right] \, 
f_1(p^2) \Bigg{\}}~. \label{ff2}
\eea
Here $\epsilon_\mu$ and $q_\mu$ are the four vector polarization and four
momentum of the photon, respectively .
To calculate the matrix elements  $ \la \gamma \ve \bar s ( 1\pm\gamma_5) 
b \ve B \ra $, we multiply both sides of  eq. (\ref{ff1}) by $p_{\mu}$ and 
use the equations of motion. However, neglecting the mass of the strange 
quark they vanish,
\beq
\la \gamma \ve \bar s ( 1\pm \gamma_5) b \ve B \ra =0
\eeq 
Substituting Eqs. (\ref{ff1}) and (\ref{ff2}) in (\ref{M1eq1}), for the 
matrix element ${\cal M}_1$ (structure dependent part) we get
\bea
{\cal M}_1 & = & \frac{\alpha G_F}{2 \sqrt{2} \, \pi} V_{tb} V_{ts}^* e 
\Bigg{\{} \epsilon_{\mu \alpha \beta \sigma} \epsilon_\alpha^* p_\beta
q_\sigma \left[ A \, \bar \tau \gamma_\mu \tau + C \, \bar \tau \gamma_\mu 
\gamma_5 \tau \right]  +  i \left[ \epsilon_\mu^* (p q) - (\epsilon^* p ) 
q_\mu \right] \left[ B \bar \tau \gamma_\mu \tau + D   \bar \tau \gamma_\mu 
\gamma_5 \tau \right]  \Bigg{\}} \nnb \\ & & 
\eea
where 
\bea
A &=& \frac{1}{m_B^2} \left[ C_9^{eff} \,g(p^2) - 2 C_7 \frac{m_b}{p^2} \, 
g_1(p^2) \right]~~, ~~
B = \frac{1}{m_B^2} \left[ C_9^{eff} \,f(p^2) - 2 C_7 \frac{m_b}{p^2} \, 
f_1(p^2) \right]~~, \nnb \\
C &=& \frac{C_{10}}{m_B^2} \,g(p^2)~~,~~D = \frac{C_{10}}{m_B^2} \,f(p^2)~~.
\eea

Note that  the neutral Higgs exchange interactions do not give any 
contribution when photon is attached to the either one of the initial quark 
lines. However, when a photon is radiated from the final $\tau$-leptons 
the situation is different and the corresponding matrix element 
(Bremsstrahlung part) is 
\bea
{\cal M}_2 & = & \frac{\alpha G_F}{2 \sqrt{2} \, \pi} V_{tb} V_{ts}^* e
i f_B    \Bigg{\{}  \left(2 m_\tau C_{10} +\frac{m_{B}^{2}}{m_{b}} 
C_{Q_{2}} \right)  \left[ \bar \tau \left(
\frac{\not\!\epsilon \! \not\!P_B}{2 p_1 q} - 
\frac{\not\!P_B \! \not\!\epsilon}{2 p_2  q} \right) \gamma_5 \tau \right] 
\nnb \\
& + & \frac{m_{B}^{2}}{m_{b}}C_{Q_{1}}\left[2~m_{\tau}\left(
\frac{1}{2p_{1}q}+\frac{1}{2p_{2}q} \right) \bar \tau \not\!\epsilon \tau 
+ \bar \tau \left(
\frac{\not\!\epsilon \! \not\!P_B}{2 p_1 q} - 
\frac{\not\!P_B \! \not\!\epsilon}{2 p_2  q} \right) \tau \right] \Bigg{\}} 
\label{M2}
\eea
where  we have used 
\bea
\la 0 \ve \bar s \gamma_\mu \gamma_5 b \ve B \ra &=& -~i f_B P_{B\mu}~, 
\nnb \\
\la 0 \ve \bar s \sigma_{\mu\nu} (1\pm\gamma_5) b \ve B \ra &=& 0~, \nnb \\
\la 0 \ve \bar s \gamma_5 b \ve B \ra &=& i f_B \frac{m_{B}^{2}}{m_{b}}~, 
\nnb \\ \la 0 \ve \bar s  b \ve B \ra &=& 0~.
\eea
and the conservation of the vector current. Here  $P_{B}$ is the momentum 
of the B-meson.

Finally, we get the total matrix element for the $B \rar \tau^+ \tau^- \gamma$ 
decay as 
\bea
{\cal M} = {\cal M}_1 + {\cal M}_2~.
\eea
To calculate the decay rate, we need the square of this matrix element. By  
summing  over the spins of the $\tau$--leptons and the polarization of the 
photon, we obtain
\bea
\ve {\cal M} \ve^2 = \ve {\cal M}_1 \ve^2 + \ve {\cal M}_2 \ve^2 + 
2 \,{\rm Re} \left( {\cal M}_1 {\cal M}_2^*\right)~,
\eea
where
\bea
\vel {\cal M}_1 \ver^2 &=& \vel \frac{\alpha G_F}{2 \sqrt{2} \, \pi} 
V_{tb} V_{ts}^* \ver^2 4 \pi \alpha \, 
\Bigg{\{} 8 \, {\rm Re} \ga B^* C + A^* D \dr p^2 \ga p_1 q - p_2 q \dr
\ga p_1 q + p_2 q \dr ~ \nnb \\
&& +~ 4 \left[ \vel C \ver^2 + \vel D \ver^2 \right] 
\left[ \ga p^2 - 2 m_\tau^2 \dr \ga \ga p_1 q \dr^2 + \ga p_2 q \dr^2
\dr - 4 m_\tau^2 \ga p_1 q \dr \ga p_2 q \dr \right]~  \nnb \\
&& +~ 4 \left[ \vel A \ver^2 + \vel B \ver^2 \right]
\Big[ \ga p^2 + 2 m_\tau^2 \dr \ga \ga p_1 q \dr^2 + \ga p_2 q \dr^2
\dr + ~ 4 m_\tau^2 \ga p_1 q \dr \ga p_2 q \dr \Big] \Bigg{\}}~, \nnb \\ & & \\ 
2\, {\rm Re} \ga {\cal M}_1 {\cal M}_2^* \dr &=&
~\vel \frac{\alpha G_F}{2 \sqrt{2} \, \pi} V_{tb} V_{ts}^* \ver^2 4 \pi \alpha \,
\Bigg{\{} 16 \, C_{10} f_B m_\tau^2 \Bigg[ {\rm Re}(A) \, 
\frac{ \ga p_1 q +  p_2 q \dr^3}{\ga p_1 q \dr \ga p_2 q \dr}~ \nnb \\
&& +~ {\rm Re}(D) \,
\frac{ \ga p_1 q +  p_2 q \dr^2  \ga p_1 q -  p_2 q \dr}
{\ga p_1 q \dr \ga p_2 q \dr} \Bigg]   \nnb \\ 
&&-~\frac{m_{B}^{2}}{m_{b}} C_{Q_{1}}\Bigg[ {\rm Re}(B) \, 
\frac{ \ga p_1 q +  p_2 q \dr^3}{\ga p_1 q \dr \ga p_2 q \dr} -{\rm Re}(C) \,
\frac{ \ga p_1 q +  p_2 q \dr^2  \ga p_1 q -  p_2 q \dr}
{\ga p_1 q \dr \ga p_2 q \dr} \Bigg] \nnb \\ 
&&+~\frac{m_{B}^{2}}{m_{b}} {\rm Re}(B) \Big[ \frac{(m_{\tau}^{2}-3 p_{2}q)
(p_{1}q)}{p_{2}q}+\frac{(2m_{\tau}^{2}- p^{2})(p_{2}q)}{p_{1}q}\Big] \Big{\}}~,
\eea
\bea
\lefteqn{\vel {\cal M}_2 \ver^2 = -~\vel \frac{\alpha G_F}{2 \sqrt{2} \, \pi}
V_{tb} V_{ts}^* \ver^2 4 \pi \alpha } \nnb \\
& & \Bigg{\{} - 16 \Big[ \left(2 m_{\tau} C_{10}+\frac{m_{B}^{2}}{m_b} 
C_{Q_{2}} \right)^{2}+
\left(\frac{m_{B}^{2}C_{Q_{1}}}{m_{b}}\right)^{2}\Big]  \nnb \\
&&~+ \frac{2m_\tau^2}{\ga p_1 q \dr^2} \Big[\left(2 m_{\tau} C_{10}+
\frac{m_{B}^{2}}{m_b} C_{Q_{2}}\right)^{2} \ga p^2 + 2 p_2 q \dr +
\left(\frac{m_{B}^{2}C_{Q_{1}}}{m_{b}}\right)^{2}
\ga p^2 + 2 p_2 q -4m_{\tau}^{2}\dr \Big] \nnb \\
&& +~ \frac{4}{p_1 q}   \Big[ \left(2 m_{\tau} C_{10}+\frac{m_{B}^{2}}{m_b} 
C_{Q_{2}}\right)^{2}+
\left(\frac{m_{B}^{2}C_{Q_{1}}}{m_{b}}\right)^{2}\Big]                   
\left[ 3 m_\tau^2 - p^2 -2 p_2 q \right] \nnb  \\
&& +~ \frac{2m_\tau^2}{\ga p_2 q \dr^2} \Big[\left(2 m_{\tau} C_{10}+
\frac{m_{B}^{2}}{m_b} C_{Q_{2}}\right)^{2} \ga p^2 + 2 p_1 q \dr +\left
(\frac{m_{B}^{2}
C_{Q_{1}}}{m_{b}}\right)^{2}
\ga p^2 + 2 p_1 q -4m_{\tau}^{2}\dr \Big]~ \nnb \\
&& +~\frac{4}{p_2 q}   \Big[ \left(2 m_{\tau} C_{10}+\frac{m_{B}^{2}}{m_b} 
C_{Q_{2}}\right)^{2}+
\left(\frac{m_{B}^{2}C_{Q_{1}}}{m_{b}}\right)^{2}\Big]                   
\left[ 3 m_\tau^2 - p^2 -2 p_1 q \right]  \nnb \\
&& +\frac{2}{\ga p_1 q \dr \ga p_2 q \dr} \Big[\left(2 m_{\tau} C_{10}+
\frac{m_{B}^{2}}{m_b} C_{Q_{2}}\right)^{2} p^2 \ga 2m_{\tau}^{2}-p^2 \dr -
\left(\frac{m_{B}^{2}C_{Q_{1}}}{m_{b}}\right)^{2}
\ga p^2 + 2 p_2 q -4m_{\tau}^{2}\dr \Big] \Big{\}} ~. \nnb \\
& &
\eea
Here $p_1,~p_2$ are momenta of the final $\tau$--leptons.

In the rest frame of the $B$--meson, the photon energy
$E_\gamma$ and the lepton energy $E_1$ are restricted in the region  given by 
\bea
0 \leq E_\gamma \leq \frac{m_B^2 - 4 m_\tau^2}{2 m_B}~, \nnb
\eea
\bea
\frac{m_B - E_\gamma}{2} - \frac{E_\gamma}{2} 
\sqrt{1 - \frac{4 m_\tau^2}{m_B^2 - 2 m_B E_\gamma}} \leq E_1 \leq
\frac{m_B - E_\gamma}{2} + \frac{E_\gamma}{2} 
\sqrt{1 - \frac{4 m_\tau^2}{m_B^2 - 2 m_B E_\gamma}}~.
\eea

In  $\vel {\cal M}_2 \ver^2$ it appears an infrared
divergence, which originates in the Bremstrahlung processes when photon is soft 
and in this limit, the $B \rar \tau^+ \tau^- \gamma$ decay cannot be
distinguished from $B \rar \tau^+ \tau^-$. Therefore, in order to cancel the 
infrared divergences in the decay rate  both processes must be considered 
together. In
ref. \cite{Aliev2} it has been   shown that infrared singular terms in 
$\vel {\cal M}_2 \ver^2$
exactly cancel the $O(\alpha)$ virtual correction in $B \rar \tau^+
\tau^-$ amplitude.  However, in this work we consider the photon in $B \rar 
\tau^+
\tau^- \gamma$ as a hard photon, following the approach described in ref. 
\cite{Aliev2} and impose a cut  
on the photon energy. The lower limit of this cut is choosen so that the 
radiated photon  can be detected in the experiments, namely $E_{\gamma }
\geq 50$ MeV ($\simeq a~m_B$ with $a \geq 0.01$).  
After integrating over the phase space and taking into account the cut for
the photon energy  we get for the decay rate
\bea
\Gamma &=& \vel \frac{\alpha G_F}{2 \sqrt{2} \, \pi} V_{tb} V_{ts}^* \ver^2 \,
\frac{\alpha}{\ga 2 \, \pi \dr^3}\, m_B^5 \pi \nnb \\
&\times& \Bigg{\{} \frac{ m_B^2 }{12} \, \int_\delta^{1-4 r} x^3 \,dx \,
\sqrt{1-\frac{4 r}{1 - x}} \, \Big[ \ga \vel A \ver^2 + \vel B \ver^2
\dr \ga 1- x+ 2 r \dr + \ga \vel C \ver^2 + \vel D \ver^2 \dr \ga 1- x - 4 r \dr 
\Big] \nnb \\
&-& 4 f_B r \Big{\{} \left(C_{10}+\frac{m_B^2}{2m_b m_\tau} C_{Q_{2}}\right) 
\int_\delta^{1-4 r} x^2 \,dx \, {\rm Re} \ga A \dr {\rm ln}
 \displaystyle{\frac{1 + \sqrt{1-\displaystyle{\frac{4 r}{1 - x}}}}
{1 -  \sqrt{1-\displaystyle{\frac{4 r}{1 - x}}}}} \nnb \\
&+& \frac{m_{B}^{2}}{2m_{b} m_{\tau}} C_{Q_{1}} \int_\delta^{1-4 r} x \,dx \, 
{\rm Re} \ga B \dr
\Big[\ga 1-x \dr \, \sqrt{1-\frac{4 r}{1 - x}}\, +(x-2 r)\, 
\displaystyle{\frac{1 + \sqrt{1-\displaystyle{\frac{4 r}{1 - x}}}}
{1 -  \sqrt{1-\displaystyle{\frac{4 r}{1 - x}}}}}\Big] \Big{\}} \nnb \\
&-& 8 f_B^2 r \, \frac{1}{m_B^2} \,\Big{\{} \left(C_{10}+\frac{m_{B}^{2}}{2m_{b }
m_{\tau}} C_{Q_{2}}\right)^2
\int_\delta^{1-4 r} dx \Big[\frac{ \ga 1-x \dr}{x} \, \sqrt{1-\frac{4 r}{1 - x}}
 \nnb \\ &+&  \ga 1 + \frac{2 r }{x} -\frac{1}{x} -x \dr\,{\rm ln}
\displaystyle{\frac{1 + \sqrt{1-\displaystyle{\frac{4 r}{1 - x}}}}
{1 -  \sqrt{1-\displaystyle{\frac{4 r}{1 - x}}}}}\nnb \\
&-&\frac{1}{r}~\left(\frac{m_{B} C_{Q_{1}}}{2m_{b}}\right)^2
\int_\delta^{1-4 r} dx \Big[(4r-1)\frac{ \ga 1-x \dr}{x} \, 
\sqrt{1-\frac{4 r}{1 - x}}\nnb \\
& + & \ga -1 + \frac{8 r^2 }{x} +\frac{1}{x} +x +\frac{r}{x}(4x-6)\dr\,{\rm ln}
\displaystyle{\frac{1 + \sqrt{1-\displaystyle{\frac{4 r}{1 - x}}}}
{1 -  \sqrt{1-\displaystyle{\frac{4 r}{1 - x}}}}}\, \Bigg] \Bigg{\}}~, 
\label{Gama1}
\eea
where $r=\displaystyle{\frac{m_\tau^2}{m_B^2}}$, $\delta=2 a$ and 
$x=\displaystyle{\frac{2 E_\gamma}{m_B}}$ is the dimensionless photon energy 
satisfying
\bea
\delta \leq x \leq 1 - \frac{4 m_\tau^2}{m_B^2}~. \nnb
\eea
In our numerical calculations,  we use the dipole forms of the form factors 
given by 
\bea
g_(p^2) &=& \frac{1~GeV}{(1-\displaystyle{\frac{p^2}{5.6^2})^2}}~,
~~~~~~~~~~
f_(p^2) = \frac{0.8~GeV}{(1-\displaystyle{\frac{p^2}{6.5^2})^2}}~,
\nnb \\ \nnb  \\ \nnb \\ 
g_1(p^2) &=& \frac{3.74~GeV^2}{(1-\displaystyle{\frac{p^2}{40.5})^2}}~, 
~~~~~~~~~
f_1(p^2) = \frac{0.68~GeV^2}{(1-\displaystyle{\frac{p^2}{30})^2}}~,
\eea
which were calculated in the framework of the light-cone QCD sum rules 
\cite{Eilam2, Buchalla2}.
\section{Results and Discussion}
In the 2HDM there are number of free parameters, namely  masses of the 
charged and neutral Higgs bosons ($m_{H^\pm}$,
$m_{h^0}$, $m_{A^0}$),  the ratio of vacuum expectation values of Higgs 
bosons,  $\tan \beta =v_2/v_1$, and the angle $\alpha$ due to the  mixing 
of neutral Higgs bosons, $A^0$ and $h^0$. The values of these parameters 
are  restricted by using the existing experimental data. The non-observation 
of charged $H^\pm$ pair in Z decays \cite{Abreu} gives the model independent 
lower bound of the mass of the charged Higgs $H^\pm$, $m_{H^\pm} \geq 44$ GeV. 
However there is no experimental upper bound for $m_{H^\pm}$ except $m_{H^\pm}
\leq 1$ TeV coming from the unitarity condition \cite{Veltman}. Further, 
top decays give $m_{H^\pm}  \geq 147$ GeV for large $\tan \beta$ \cite{Abe}. 
The other parameter of 2HDM, $\tan \beta$, is restricted as $\tan \beta >0.7 $ 
from $Z\rightarrow \bar b \, b$ decay \cite{Grant}. The  ratio $\tan \beta 
/m_{H^\pm}$ can also be restricted and it has been estimated as $\tan \beta 
/m_{H^\pm} \leq 0.38$ GeV$^{-1}$ \cite{Acciari} and 
$\tan \beta /m_{H^\pm} \leq 0.46$ GeV$^{-1}$ \cite{ALEPH} from the experimental 
results of the branching ratios of the decays $B\rightarrow \tau \, \bar{\nu}$ 
and $B\rightarrow X\, \tau \, \bar{\nu}$. The upper bound has also been given 
for the same ratio as $\tan \beta /m_{H^\pm} =0.06$ GeV$^{-1}$in the case that 
sufficient data could be taken and the theoretical uncertainties  could be 
reduced  for the exclusive decay $B\rightarrow D \, \tau \bar{\nu}$ \cite{Kiers}.
 Recently, the relation between $m_{H^\pm}$ and $\tan \beta $ has been estimated 
in \cite{Aliev3} , taking into account the CLEO measurement of the decay 
$B\rightarrow X_s \gamma$
\cite{Alam}, 
\begin{eqnarray}
Br (B\rightarrow X_s\gamma)= (3.15\pm 0.35\pm 0.32)\, 10^{-4} \,\, ,
\label{br2}
\end{eqnarray}

In our calculations, we take the masses $m_{h^0}$ and $m_{A^0}$ equal and not 
too heavy  since the b-quark dipole moment is strongly sensitive to the difference
between these masses in the 2HDM \cite{Iltan}. Further, we choose the value of 
the angle $\alpha$ as being zero since the mixing between $h^{0}$ and $A^{0}$ is 
weak. For completeness, we have also checked the dependence of the branching 
ratio on  $\alpha$ for the fixed values the other 2HDM parameters and seen  that  
this dependence is negligible.

In the present work, we study the 2HDM parameters dependence of the BR and 
dimensionless photon energy dependence of the differential branching ratio 
(dBR/dx) in Model I and II. Doing this, we have used the input parameters 
given in Table I. 

In Fig.1, we present $dBR(B\rightarrow \tau^{+} \tau^{-} \gamma )/dx$ as a 
function of  $x=2E_{\gamma}/m_{B}$ in the  SM and in Model II  for 
$m_{H^\pm}=400$ GeV and $\tan \beta =2$. We do not display the predictions 
of Model I there,  since they are very close to those  of Model II. In this 
figure, curves with sharp peaks represent the long distance contributions. 
From Fig. 1,  we see that there is an enhancement in the 2HDM compared to the 
SM case.

Fig 2. shows the dependence of the BR on the Higgs boson mass $m_{H^\pm}$  for 
different values of the parameter $\tan\beta$ for Model I and II, as well as 
for  the SM. We again observe an enhancement for the BR in 2HDM compared to the 
SM case. For example, for $m_{H^\pm}=400~GeV$ and $\tan\beta =2$ , 
$BR(B\rightarrow \tau^{+}\tau^{-}\gamma )=4.18 \times 10^{-8}$ in Model I, and 
$BR(B\rightarrow \tau^{+}\tau^{-}\gamma )=4.20 \times 10^{-8}$ in Model II. 
These values are greater than the SM predictions, which is 
$BR(B\rightarrow \tau^{+}\tau^{-}\gamma )=4.13 \times 10^{-8}$.
In addition, the $m_{H^\pm}$ dependence of the BR becomes weaker  with increasing
values of  $\tan \beta$ for both models.

We present the BR as a function of $\tan \beta $ for different values of 
$m_{H^\pm}$ in Model I and II in Fig. 3 and Fig. 4, respectively. It is seen 
that additional contributions coming from neutral Higgs exchange diagrams 
(i.e., contributions with $C_{Q_i}\neq 0$ ) causes the BR to increase with the 
increasing values of $\tan \beta$ in contrast to the case that  neutral Higgs do 
not contribute ($C_{Q_i}=0$). The reason for these two different behaviors can 
easily be understood by comparing eqs.(\ref{CoeffW}) and (\ref{CH}) , which 
represent the neutral Higgs bosons and the remaining contributions, respectively, 
namely  the first one is proportional to $\tan^2 \beta$ , while the second is  
$1/\tan ^2\beta $ so that  for the larger values of $\tan \beta$ , neutral Higgs 
contributions dominate in the BR.

As a conclusion, we observe an enhancement in the differential branching ratio 
and the branching ratio of the exclusive process $B\rightarrow \tau^+ \tau^- 
\gamma $ in the framework of the 2HDM as compared to the SM. Further, this 
enhancement becomes more detectable for large $\tan \beta $ values lying in 
experimentally restricted regions. Therefore, the measurement of this exclusive 
decay gives important clues about the  new physics beyond the SM, corresponding   
model parameters and also the effects of neutral Higgs contributions.
\newpage
{\bf \LARGE {Appendix}} \\
\begin{appendix}
\section{The operator basis }
The operator basis in the  2HDM (Model I and II) for the process under 
consideration  is \cite{Grinstein,Misiak}
\begin{eqnarray}
 O_1 &=& (\bar{s}_{L \alpha} \gamma_\mu c_{L \beta})
               (\bar{c}_{L \beta} \gamma^\mu b_{L \alpha}), \nonumber   \\
 O_2 &=& (\bar{s}_{L \alpha} \gamma_\mu c_{L \alpha})
               (\bar{c}_{L \beta} \gamma^\mu b_{L \beta}),  \nonumber   \\
 O_3 &=& (\bar{s}_{L \alpha} \gamma_\mu b_{L \alpha})
               \sum_{q=u,d,s,c,b}
               (\bar{q}_{L \beta} \gamma^\mu q_{L \beta}),  \nonumber   \\
 O_4 &=& (\bar{s}_{L \alpha} \gamma_\mu b_{L \beta})
                \sum_{q=u,d,s,c,b}
               (\bar{q}_{L \beta} \gamma^\mu q_{L \alpha}),   \nonumber  \\
 O_5 &=& (\bar{s}_{L \alpha} \gamma_\mu b_{L \alpha})
               \sum_{q=u,d,s,c,b}
               (\bar{q}_{R \beta} \gamma^\mu q_{R \beta}),   \nonumber  \\
 O_6 &=& (\bar{s}_{L \alpha} \gamma_\mu b_{L \beta})
                \sum_{q=u,d,s,c,b}
               (\bar{q}_{R \beta} \gamma^\mu q_{R \alpha}),  \nonumber   \\  
 O_7 &=& \frac{e}{16 \pi^2}
          \bar{s}_{\alpha} \sigma_{\mu \nu} (m_b R + m_s L) b_{\alpha}
                {\cal{F}}^{\mu \nu},                             \nonumber  \\
 O_8 &=& \frac{g}{16 \pi^2}
    \bar{s}_{\alpha} T_{\alpha \beta}^a \sigma_{\mu \nu} (m_b R +
m_s L)  
          b_{\beta} {\cal{G}}^{a \mu \nu} \nonumber \,\, , \\  
 O_9 &=& \frac{e}{16 \pi^2}
          (\bar{s}_{L \alpha} \gamma_\mu b_{L \alpha})
              (\bar{l} \gamma^\mu l)  \,\, ,    \nonumber    \\
 O_{10} &=& \frac{e}{16 \pi^2}
          (\bar{s}_{L \alpha} \gamma_\mu b_{L \alpha})
              (\bar{l} \gamma^\mu \gamma_{5} l)  \,\, ,    \nonumber  \\
Q_1&=&   \frac{e^2}{16 \pi^2}(\bar{s}^{\alpha}_{L}\,b^{\alpha}_{R})\,
(\bar{\tau}\tau ) \nnb  \\ 
Q_2&=&    \frac{e^2}{16 \pi^2}(\bar{s}^{\alpha}_{L}\,b^{\alpha}_{R})\,
(\bar{\tau} \gamma_5 \tau ) \nnb \\
Q_3&=&    \frac{g^2}{16 \pi^2}(\bar{s}^{\alpha}_{L}\,b^{\alpha}_{R})\,
\sum_{q=u,d,s,c,b }(\bar{q}^{\beta}_{L} \, q^{\beta}_{R} ) \nnb \\
Q_4&=&  \frac{g^2}{16 \pi^2}(\bar{s}^{\alpha}_{L}\,b^{\alpha}_{R})\,
\sum_{q=u,d,s,c,b } (\bar{q}^{\beta}_{R} \, q^{\beta}_{L} ) \nnb \\
Q_5&=&   \frac{g^2}{16 \pi^2}(\bar{s}^{\alpha}_{L}\,b^{\beta}_{R})\,
\sum_{q=u,d,s,c,b } (\bar{q}^{\beta}_{L} \, q^{\alpha}_{R} ) \nnb \\
Q_6&=&   \frac{g^2}{16 \pi^2}(\bar{s}^{\alpha}_{L}\,b^{\beta}_{R})\,
\sum_{q=u,d,s,c,b } (\bar{q}^{\beta}_{R} \, q^{\alpha}_{L} ) \nnb \\
Q_7&=&   \frac{g^2}{16 \pi^2}(\bar{s}^{\alpha}_{L}\,\sigma^{\mu \nu} \, 
b^{\alpha}_{R})\,
\sum_{q=u,d,s,c,b } (\bar{q}^{\beta}_{L} \, \sigma_{\mu \nu } 
q^{\beta}_{R} ) \nnb \\
Q_8&=&    \frac{g^2}{16 \pi^2}(\bar{s}^{\alpha}_{L}\,\sigma^{\mu \nu} \, 
b^{\alpha}_{R})\,
\sum_{q=u,d,s,c,b } (\bar{q}^{\beta}_{R} \, \sigma_{\mu \nu } q^{\beta}_{L} ) 
\nnb \\ 
Q_9&=&   \frac{g^2}{16 \pi^2}(\bar{s}^{\alpha}_{L}\,\sigma^{\mu \nu} \, 
b^{\beta}_{R})\,
\sum_{q=u,d,s,c,b }(\bar{q}^{\beta}_{L} \, \sigma_{\mu \nu } q^{\alpha}_{R} )
\nnb \\
Q_{10}&= & \frac{g^2}{16 \pi^2}(\bar{s}^{\alpha}_{L}\,\sigma^{\mu \nu} \, 
b^{\beta}_{R})\,
\sum_{q=u,d,s,c,b }(\bar{q}^{\beta}_{R} \, \sigma_{\mu \nu } q^{\alpha}_{L} ) 
\label{op1}
\end{eqnarray}
where  
$\alpha$ and $\beta$ are $SU(3)$ colour indices and
${\cal{F}}^{\mu \nu}$ and ${\cal{G}}^{\mu \nu}$
are the field strength tensors of the electromagnetic and strong
interactions, respectively.
\section{The Initial values of the Wilson coefficients.}
The initial values of the Wilson coefficients for the relevant process 
in the SM are \cite{Grinstein2}
\begin{eqnarray}
C^{SM}_{1,3,\dots 6,11,12}(m_W)&=&0 \nonumber \, \, , \\
C^{SM}_2(m_W)&=&1 \nonumber \, \, , \\
C_7^{SM}(m_W)&=&\frac{3 x^3-2 x^2}{4(x-1)^4} \ln x+
\frac{-8x^3-5 x^2+7 x}{24 (x-1)^3} \nonumber \, \, , \\
C_8^{SM}(m_W)&=&-\frac{3 x^2}{4(x-1)^4} \ln x+
\frac{-x^3+5 x^2+2 x}{8 (x-1)^3}\nonumber \, \, , \\ 
C_9^{SM}(m_W)&=&-\frac{1}{sin^2\theta_{W}} B(x) +
\frac{1-4 \sin^2 \theta_W}{\sin^2 \theta_W} C(x)-D(x)+\frac{4}{9},
\nonumber \, \, , \\
C_{10}^{SM}(m_W)&=&\frac{1}{\sin^2\theta_W}
(B(x)-C(x))\nonumber \,\, , \\
C_{Q_i}^{SM}(m_W) & = & 0~~~ i=1,..,10~.
\end{eqnarray}
The initial values for the additional part due to charged Higgs bosons are 
\begin{eqnarray}
C^{H}_{1,\dots 6 }(m_W)&=&0 \nonumber \, , \\
C_7^{H}(m_W)&=& X \, F_{1}(y)\, + \, Y \,  F_{2}(y) \nonumber  \, \, , \\
C_8^{H}(m_W)&=& X \,  G_{1}(y) \, + \, Y \, G_{2}(y) \nonumber\, \, , \\
C_9^{H}(m_W)&=&  X \,  H_{1}(y) \nonumber  \, \, , \\
C_{10}^{H}(m_W)&=& X \,  L_{1}(y)  \label{CH} \, \, , 
\end{eqnarray}
and due to the neutral  Higgs bosons are \cite{Dai}
\begin{eqnarray}
C_{Q_1}^{H}(m_W)&=& \frac{m_b m_{\tau}}{m_{h^0}^2} 
\frac{1}{\sin^2\theta_W} \frac{x}{4} \, X^{-1} \, \Big{\{} 
\frac{\sin^2 2\alpha}{2m^2_{H^{\pm}}} \left( m^2_{h^0}-
\frac{(m^2_{h^0}-m^2_{H^0})^2}{2 m^2_{H^0}}\right) f_3(y) \nnb \\
& + & (\sin^2\alpha +h \cos^2\alpha ) f_1(x,y)+[m^2_{h^0}/m^2_W+ 
(\sin^2\alpha +h \cos^2\alpha )(1-z)]f_2(x,y) \Big{\}}  \nnb \\
C_{Q_2}^{H}(m_W)&=& -\frac{m_b m_{\tau}}{m_{A^0}^2} 
\,X^{-1}\, \Big{\{} f_1(x,y)+
\left( 1+\frac{(m^2_{H^{\pm}}-m^2_{A^0})}{2 m^2_{W}}\right)
f_2(x,y) \Big{\}} \nnb \\
C_{Q_3}^{H}(m_W)&=& \frac{m_b e^2}{m_{\tau} g^2} 
(C_{Q_{1}}(m_W)+ C_{Q_{2}}(m_W)) \nnb \\
C_{Q_4}^{H}(m_W)&=& \frac{m_b e^2}{m_{\tau} g^2}
(C_{Q_{1}}(m_W)-C_{Q_{2}}(m_W))\nnb \\
C_{Q_i}(m_W)& = & 0~, i=5,..,10 ~, \label{CoeffW}
\end{eqnarray}
where
\begin{eqnarray}
& & x=\frac{m_t^2}{m_W^2}~~~,~~~y=
\frac{m_t^2}{m_H^{\pm}}~~~,~~~z=\frac{x}{y}~~~,~~~
h=\frac{m_{h^0}^2}{m_{H^0}^2}~~~,~~~
 f_1(x,y)=\frac{x\ln x }{x-1}-\frac{y\ln y}{y-1}~,\nnb \\ 
& &  f_2(x,y)=\frac{x\ln y }{(z-x)(x-1)}+
\frac{ \ln z}{(z-1)(x-1)}~,~f_3(y)=\frac{1-y+y\ln y}{(y-1)^2}~.
\label{CoeffH}
\end{eqnarray}
and
\beq
X=\frac{1}{\tan^2\beta}~~~\left(\frac{1}{\tan^2\beta} \right) ~~,~~
Y  = \frac{-1}{\tan^2\beta} ~~~~(1) ~~~~in~~Model~I~(II)    
\eeq

The explicit forms of the functions $F_{1(2)}(y)$, $G_{1(2)}(y)$, 
$H_{1}(y)$ and $L_{1}(y)$ are given as
\begin{eqnarray}
F_{1}(y)&=& \frac{y(7-5y-8y^2)}{72 (y-1)^3}+\frac{y^2 (3y-2)}{12(y-1)^4}
\,\ln y \nonumber  \,\, , \\ 
F_{2}(y)&=& \frac{y(5y-3)}{12 (y-1)^2}+\frac{y(-3y+2)}{6(y-1)^3}\, \ln y 
\nonumber  \,\, ,\\ 
G_{1}(y)&=& \frac{y(-y^2+5y+2)}{24 (y-1)^3}+\frac{-y^2} {4(y-1)^4} \, \ln y
\nonumber  \,\, ,\\ 
G_{2}(y)&=& \frac{y(y-3)}{4 (y-1)^2}+\frac{y} {2(y-1)^3} \, \ln y 
\nonumber\,\, ,\\
H_{1}(y)&=& \frac{1-4 sin^2\theta_W}{sin^2\theta_W}\,\, \frac{x
y}{8}\,\left[ 
\frac{1}{y-1}-\frac{1}{(y-1)^2} \ln y \right]-y \left[\frac{47 y^2-79 y+38}{108
(y-1)^3}-\frac{3 y^3-6 y+4}{18(y-1)^4} \ln y \right] 
\nonumber  \,\, , \\ 
L_{1}(y)&=& \frac{1}{sin^2\theta_W} \,\,\frac{x y}{8}\, \left[-\frac{1}{y-1}+
\frac{1}{(y-1)^2} \ln y \right]
\nonumber  \,\, .\\ 
\label{F1G1}
\end{eqnarray}
Finally, the initial values of the coefficients in the 2HDM are
\begin {eqnarray}   
C_i^{2HDM}(m_{W})&=&C_i^{SM}(m_{W})+C_i^{H}(m_{W})
\label{CiW}
\end{eqnarray}
Using these initial values, we can calculate the coefficients 
$C_{i}^{2HDM}(\mu)$ and $C^{2HDM}_{Q_i}(\mu)$ 
at any lower scale in the effective theory 
with five quarks, namely $u,c,d,s,b$ similar to the SM case. Wilson 
coefficients $C_{7}^{2HDM}(\mu)$, $C_{9}^{2HDM}(\mu)$,$C_{10}^{2HDM}(\mu)$, 
$C^{2HDM}_{Q_1}(\mu )$ and $C^{2HDM}_{Q_2}(\mu )$ play the essential role 
in this process and the others enter into expressions due to operator mixing.
For completeness we would like to give the explicit expressions of the
cofficients essential in this process. 
The effective coefficient $C_{7}^{eff}(\mu)$ is defined as \cite{Iltan2}
\begin{eqnarray}
C_{7}^{eff}(\mu)&=&C_{7}^{2HDM}(\mu)+ Q_d \, 
(C_{5}^{2HDM}(\mu) + N_c \, C_{6}^{2HDM}(\mu))\nonumber \, \, , \\
&+& Q_u\, (\frac{m_c}{m_b}\, C_{12}^{2HDM}(\mu) + N_c \, 
\frac{m_c}{m_b}\,C_{11}^{2HDM}(\mu)) \, \, ,
\label{C7eff}
\end{eqnarray}
where the leading order QCD corrected Wilson coefficients 
$C_{7}^{LO, 2HDM}(\mu)$ are given by \cite{Grinstein,Misiak,Buras}:
\begin{eqnarray} 
C_{7}^{LO, 2HDM}(\mu)&=& \eta^{16/23} C_{7}^{2HDM}(m_{W})+(8/3) 
(\eta^{14/23}-\eta^{16/23}) C_{8}^{2HDM}(m_{W})\nonumber \,\, \\
&+& C_{2}^{2HDM}(m_{W}) \sum_{i=1}^{8} h_{i} \eta^{a_{i}} \,\, , 
\label{LOwils}
\end{eqnarray}
and $\eta =\alpha_{s}(m_{W})/\alpha_{s}(\mu)$, $h_{i}$ and $a_{i}$ are 
the numbers which appear during the evaluation \cite{Buras}. 
The perturbative part of the Wilson coefficient $C_9^{eff}(\mu)$ 
can be defined as \cite{Misiak,Buras}:
\begin{eqnarray} 
C_9^{pert}(\mu)&=& C_9^{2HDM}(\mu) \tilde\eta (\hat s) \nonumber 
\\ &+& h(z, \hat s) \left( 3 C_1(\mu) + C_2(\mu) + 3 C_3(\mu) + 
C_4(\mu) + 3 C_5(\mu) + C_6(\mu) \right) \nonumber \\
&- & \frac{1}{2} h(1, \hat s) \left( 4 C_3(\mu) + 4 C_4(\mu) + 3
C_5(\mu) + C_6(\mu) \right) \\
&- &  \frac{1}{2} h(0, \hat s) \left( C_3(\mu) + 3 C_4(\mu) \right) +
\frac{2}{9} \left( 3 C_3(\mu) + C_4(\mu) + 3 C_5(\mu) + C_6(\mu)
\right) \nonumber \,\, .
\label{eqC9ef}
\end{eqnarray}
Here the contributions of the coefficients  
$C_1(\mu)$, ...., $C_6(\mu)$ are due to the operator mixing.
In eq. (30) $\tilde\eta(\hat s)$ represents the one gluon
correction to the matrix element $O_9$ with $m_s=0$ \cite{Misiak} and
the function $h(z,\hat s)$ arises from the one loop contributions of the
four quark operators $O_1, ... ,O_6$ .
Their explicit expressions are
\begin{eqnarray}
\tilde\eta(\hat s) = 1 + \frac{\alpha_{s}(\mu)}{\pi}\, \omega(\hat s)\,\, ,
\label{eta}
\end{eqnarray}
where
\begin{eqnarray} 
\omega(\hat s) &=& - \frac{2}{9} \pi^2 - \frac{4}{3}\mbox{Li}_2(\hat s) 
-\frac{2}{3} \ln {\hat s} \ln(1-{\hat s})-\frac{5+4{\hat s}}{3(1+2{\hat s})}
\ln(1-{\hat s}) - \nonumber \\
& &  \frac{2 {\hat s} (1+{\hat s}) (1-2{\hat s})}
{3(1-{\hat s})^2 (1+2{\hat s})} \ln {\hat s} + \frac{5+9{\hat s}-6{\hat s}^2}{6
(1-{\hat s}) (1+2{\hat s})} \,\, , 
\label{omega}
\end{eqnarray}
and 
\begin{eqnarray}
h(z, \hat s) &=& -\frac{8}{9}\ln\frac{m_b}{\mu} - \frac{8}{9}\ln z +
\frac{8}{27} + \frac{4}{9} x \\
& & - \frac{2}{9} (2+x) |1-x|^{1/2} \left\{
\begin{array}{ll}
\left( \ln\left| \frac{\sqrt{1-x} + 1}{\sqrt{1-x} - 1}\right| - 
i\pi \right), &\mbox{for } x \equiv \frac{4z^2}{\hat s} < 1 \nonumber \\
2 \arctan \frac{1}{\sqrt{x-1}}, & \mbox{for } x \equiv \frac
{4z^2}{\hat s} > 1,
\end{array}
\right. \\
h(0, \hat s) &=& \frac{8}{27} -\frac{8}{9} \ln\frac{m_b}{\mu} - 
\frac{4}{9} \ln\hat s + \frac{4}{9} i\pi \,\, , 
\label{hfunc}
\end{eqnarray}
where $z=\frac{m_c}{m_b}$ and $\hat{s}=\frac{p^2}{m_b^2}$.
In addition to the perturbative part, there exist also the long distance (LD) 
one due to the  conversion of the real $\bar{c}c$ into the lepton pair 
$\tau^+ \tau^-$, described by the reaction $B\rightarrow \gamma
\psi_i\rightarrow \gamma \tau^+ \tau^-$, where $i=1,..,6$. 
Adding this contribution to the perturbative one coming from the 
$c\bar{c}$ loop, the NLO QCD corrected $C_9^{eff}(\mu)$ can be
written as: 
\begin{eqnarray}
C_9^{eff}(\mu)=C_9^{pert}(\mu)+ Y_{reson}(\hat{s})\,\, ,
\label{C9efftot}
\end{eqnarray}
where $Y_{reson}(\hat{s})$ in NDR scheme is defined as
\begin{eqnarray}
Y_{reson}(\hat{s})&=&-\frac{3}{\alpha^2_{em}}\kappa \sum_{V_i=\psi_i}
\frac{\pi \Gamma(V_i\rightarrow \tau^+ \tau^-)m_{V_i}}{q^2-m_{V_i}+i m_{V_i}
\Gamma_{V_i}} \nonumber \\
& & \left( 3 C_1(\mu) + C_2(\mu) + 3 C_3(\mu) + 
C_4(\mu) + 3 C_5(\mu) + C_6(\mu) \right).
\label{Yres}
\end{eqnarray}
The phenomenological parameter $\kappa$ in eq. (\ref{Yres}) is taken as 
$2.3$ \cite{Ali}.

Finally, the Wilson coefficients $C_{Q_1}(\mu )$ and $C_{Q_2}(\mu )$ are given by \cite{Dai}
\beq
C_{Q_i}(\mu )=\eta^{-12/23}\,C_{Q_i}(m_W)~,~i=1,2~. 
\eeq
\end{appendix}
\newpage
\begin{table}[h]
        \begin{center}
        \begin{tabular}{|l|l|}
        \hline
        \multicolumn{1}{|c|}{Parameter} & 
                \multicolumn{1}{|c|}{Value}     \\
        \hline \hline
        $m_c$                   & $1.4$ (GeV) \\
        $m_b$                   & $4.8$ (GeV) \\
        $\alpha_{em}^{-1}$      & 137         \\
        $|V_{tb}V_{ts}^*|$            & 0.045 \\
        $m_{B_s}$             & $5.28$ (GeV) \\
        $\tau (B_s)$   & $1.64 \times 10^{-12}$ (s) \\
        $m_{t}$             & $176$ (GeV) \\
        $m_{W}$             & $80$ (GeV) \\
        $m_{Z}$             & $91.19$ (GeV) \\
        $m_{\tau}$     & $1.78$ (GeV) \\
        $m_{h^0}$    & $80$ (GeV) \\
        $m_{H^0}$  & $150$ (GeV) \\
        $m_{A^0}$   & $80$ (GeV) \\
        $\mu $   & $m_b$ \\
        $\Lambda_{QCD}$             & $0.225$ (GeV) \\
        $\alpha_{s}(m_Z)$             & $0.117$  \\
        $sin\theta_W$             & $0.2325$  \\
        \hline
        \end{tabular}
        \end{center}
\caption{The values of the input parameters used in the numerical
          calculations.}
\label{input}
\end{table}

\newpage
\begin{figure}[htb]
\vskip -3.0truein
\centering
\epsfxsize=6.8in
\leavevmode\epsffile{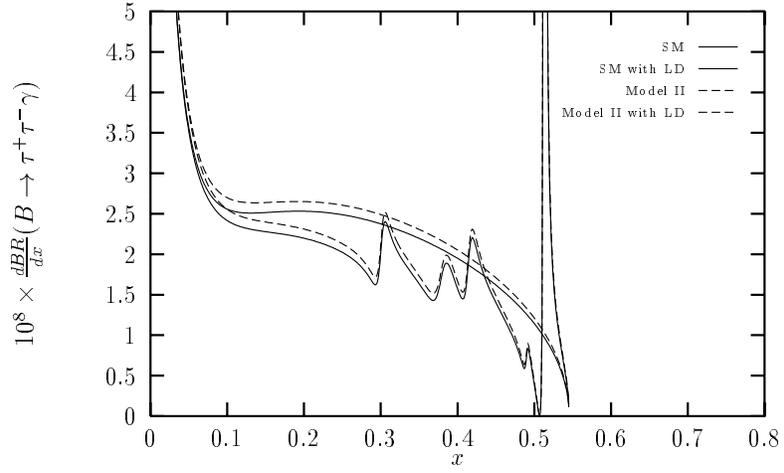}
\vskip -3.0truein
\caption[]{Differential branching ratio as a function of  $x=2E_{\gamma}/m_B$
in the SM and Model II for $m_{H^\pm}=400$ GeV and $\tan \beta =2$. In this figure, 
curves with sharp peaks represent the long distance contributions.} 
\label{dBRM2SMg2}
\end{figure}
\begin{figure}[htb]
\vskip -3.0truein
\centering
\epsfxsize=6.8in
\leavevmode\epsffile{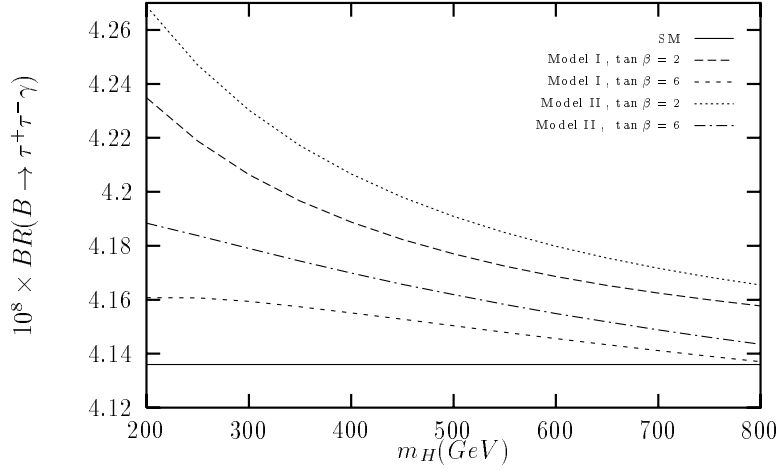}
\vskip -3.0truein
\caption[]{Branching ratio as a function of $m_{H^\pm}$ in the SM, Model I and 
II for different values of $\tan \beta$.}
\label{BRmHg3}
\end{figure}
\begin{figure}[htb]
\vskip -3.0truein
\centering
\epsfxsize=6.8in
\leavevmode\epsffile{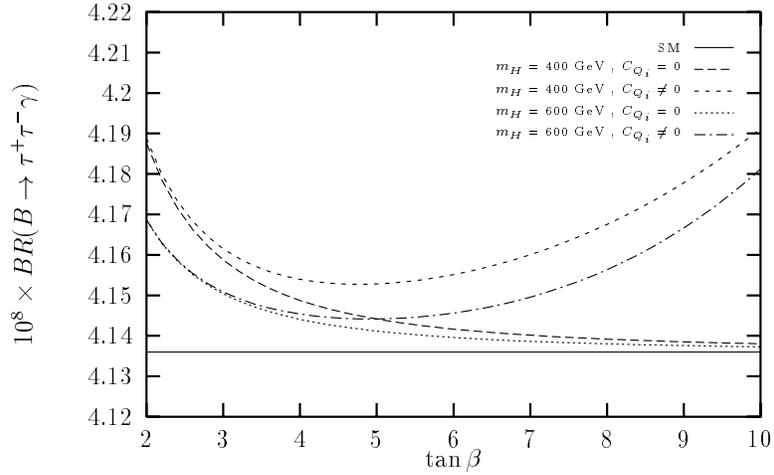}
\vskip -3.0truein
\caption[]{Branching ratio as a function of $\tan \beta$ in the SM and Model I 
for different values of $m_{H^\pm}$. Curves with $C_{Q_i}\neq 0$ ($C_{Q_i}=0 $ ) 
represent the contributions including (not including) the neutral Higgs boson 
interactions.}
\label{BRtgM1g5}
\end{figure}
\begin{figure}[htb]
\vskip -3.0truein
\centering
\epsfxsize=6.8in
\leavevmode\epsffile{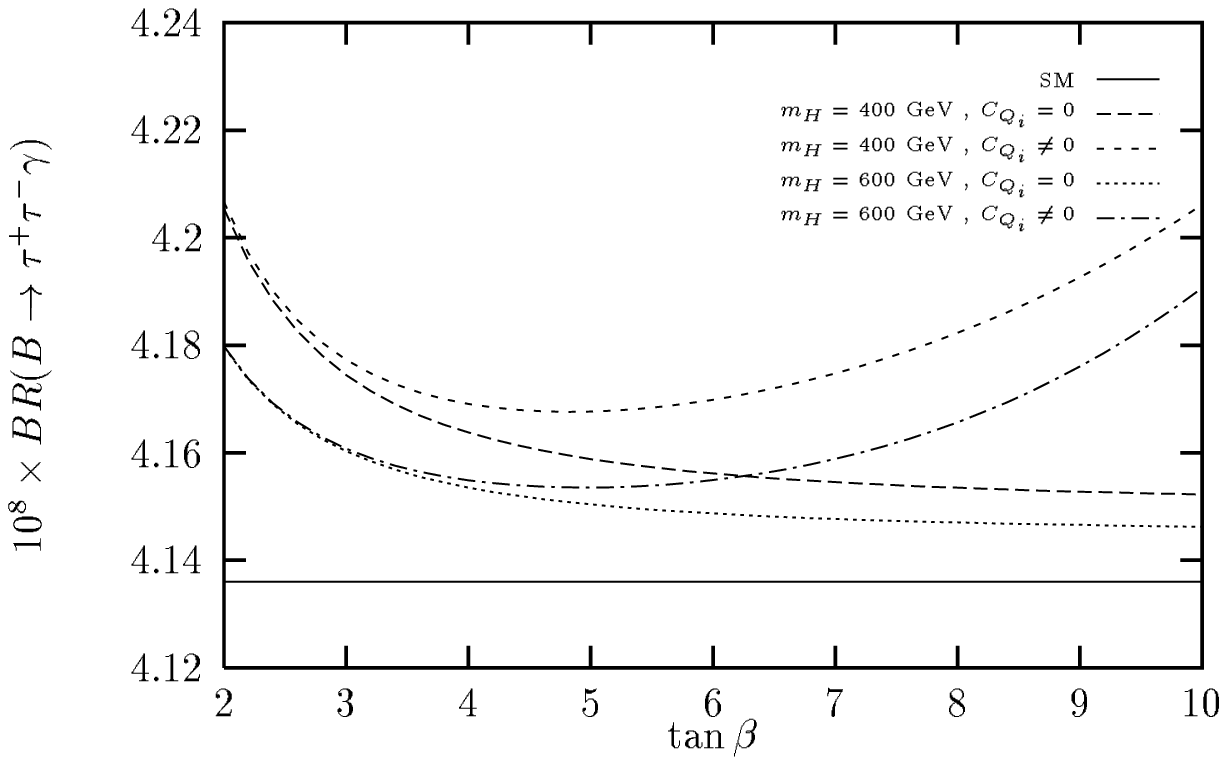}
\vskip -3.0truein
\caption[]{The same as Fig \ref{BRtgM1g5}, but for Model II.}
\label{BRtgM2g4}
\end{figure}


\newpage
\begin{thebibliography}{99}
\bibitem{Hewet} J. L. Hewet, in L. De Porcel, C. Dunwoode (Eds.), Proc. of
the 21$^{st}$ Annual SLAC Summer Institute, SLAC-PUB-6521,1994.
\bibitem{Eilam1} G. Eilam, C.-D. L\"{u} and D.-X. Zhang, {\it Phys. Lett.} 
{\bf B 391} (1997) 461.
\bibitem{Aliev1} T. M. Aliev, A. \"{O}zpineci, and M.Savc{\i}, 
{\it Phys. Rev.} {\bf D 55} (1997) 7059.
\bibitem{Aliev2} T. M. Aliev, N. K. Pak, and M.Savc{\i}, {\it Phys. Lett.} 
{\bf B 424} (1998) 175.
\bibitem{Glashow} S. Glashow and S. Weinberg, {\it Phys. Rev.} {\bf D 15} 
(1977) 1958.
\bibitem{Dai} Y.-B. Dai, C.-S. Huang and H.-W. Huang, {\it Phys. Lett.} 
{\bf B 390} (1997) 257.
\bibitem{Eilam2} G. Eilam, I. Halperin and R. R. Mendel, {\it Phys. Lett.} 
{\bf B 361} (1995) 137. \\
T.  M. Aliev, A. \"{O}zpineci and M. Savc{\i}, {\it ibid.} 
{\bf B 393} (1997) 143.
\bibitem{Buchalla2} G. Buchalla, and A. J. Buras {\it Nucl. Phys.} 
{\bf B 400} (1993) 225.
\bibitem{Abreu} P. Abreu {\it et al}.,  {\it Z. Phys.} {\bf C 64} (1994) 183 ; 
G. Alexander {\it et al}.,  {\it Phys. Lett.} {\bf B 370} (1996) 174.
\bibitem{Veltman} M. Veltman, Acta Phys. Polon. {\bf B 8} (1977) 475; 
B. W. Lee, C. Quigg, H. B. Thacker, {\it  Phys. Rev. } {\bf D 16} (1977) 253; 
M. Veltman, {\it Phys. Lett.} {\bf B 70} (1977) 253. 
\bibitem{Abe} F. Abe {\it et al}., {\it Phys. Rev. Lett.} {\bf 79} (1997) 357.
\bibitem{Grant} A. K. Grant, {\it Phys. Rev.} {\bf D 51} (1995) 207.
\bibitem{Acciari} M. Acciari, {\it et al}., 
{\it Phys. Lett.} {\bf B 396} (1997) 327.
\bibitem{ALEPH} ALEPH Collaboration, contributed to ICHEP, Warsaw, Poland, 
25-31 July 1996, Pr. No:PA10-019.
\bibitem{Kiers} K. Kiers, and A. Soni, {\it Phys. Rev.} {\bf D 56} (1997) 5786.
\bibitem{Aliev3} T. M. Aliev, G. Hiller and E. O. Iltan, {\it Nucl. Phys.} 
{\bf B 515} (1998) 321.
\bibitem{Alam} M. S. Alam, CLEO Collaboration, to appear in ICHEP98 Conference 
(1998); R. Barate {\it et al}., ALEPH Collaboration , {\it Phys. Lett.} 
{\bf B 429} (1998) 169.
\bibitem{Iltan} E. O. Iltan, hep-ph/9903433.
\bibitem{Grinstein} B. Grinstein, R. Springer and M. Wise, {\bf Nucl. Phys.} 
{\bf B 339} (1990) 269; R. Grigjanis, P. J. O'Donnell, M. Sutherland and 
H. Navelet, {\it Phys. Lett.} {\bf B 213} (1988) 355; Erratum: {\it ibid} 
{\bf B 286} (1992) 413; G. Cella, G. Curci, G. Ricciardi and A.
Vicer$\acute{e}$, {\it Phys. Lett.} {\bf B 325} (1994) 227; {\it ibid} 
{\it Nucl. Phys.} {\bf B431} (1994) 417.
\bibitem{Misiak} M. Misiak, {\it Nucl. Phys.} {\bf B 393} (1993) 23; 
Erratum: {\it ibid.} {\bf B 439} (1995) 461.
\bibitem{Grinstein2} B. Grinstein, M. J. Savage and M. Wise, {\it Nucl. Phys.} 
{\bf B 319} (1989) 271.
\bibitem{Iltan2} E. O. Iltan and T. M. Aliev, {\it Phys. Rev.} {\bf D 58}
(1998) 095014.
\bibitem{Buras} A. J. Buras and M. M\"{u}nz, {\it Phys. Rev.} {\bf D 52} 
(1995) 186.
\bibitem{Ali} A. Ali, T. Mannel and T. Morozumi, {\it Phys. Lett.} {\bf B 273} 
(1991) 505.
\end{thebibliography}
\end{document}